\documentclass[twocolumn,english,aps,pra,amsmath,amssymb,nofootinbib,showpacs,superscriptaddress]{revtex4-1}
\usepackage[scaled=0.92]{helvet}
\usepackage[T1]{fontenc}
\usepackage[latin9]{inputenc}
\usepackage{amsmath}
\usepackage{amssymb}
\usepackage{graphicx}
\usepackage{esint}

\makeatletter

\usepackage{textcomp}
\usepackage{mathrsfs}

\@ifundefined{textcolor}{}{%
 \definecolor{BLACK}{gray}{0}
 \definecolor{WHITE}{gray}{1}
 \definecolor{RED}{rgb}{1,0,0}
 \definecolor{GREEN}{rgb}{0,1,0}
 \definecolor{BLUE}{rgb}{0,0,1}
 \definecolor{CYAN}{cmyk}{1,0,0,0}
 \definecolor{MAGENTA}{cmyk}{0,1,0,0}
 \definecolor{YELLOW}{cmyk}{0,0,1,0}
}

\usepackage[english]{babel}
\usepackage{latexsym}
\usepackage{graphics}
\usepackage{epsfig}
\usepackage{color}
\usepackage{bm}

\makeatother

\usepackage{babel}
\begin{document}

\title{Collective modes of a two-dimensional Fermi gas at finite temperature}

\author{Brendan C. Mulkerin}

\affiliation{Centre for Quantum and Optical Science, Swinburne University of Technology,
Melbourne 3122, Australia.}

\author{Xia-Ji Liu}

\affiliation{Centre for Quantum and Optical Science, Swinburne University of Technology,
Melbourne 3122, Australia.}

\author{Hui Hu}

\affiliation{Centre for Quantum and Optical Science, Swinburne University of Technology,
Melbourne 3122, Australia.}

\date{\today}
\begin{abstract}
In this work we examine the breathing mode of a strongly interacting
two-dimensional Fermi gas and the role of temperature on the anomalous
breaking of scale invariance. By calculating the equation of state
with different many-body $T$-matrix theories and the virial expansion
approach, we obtain a hydrodynamic equation of the harmonically trapped
Fermi gas (with trapping frequency $\omega_{0}$) through the local
density approximation. By solving the hydrodynamic equation we determine
the breathing mode frequencies as functions of interaction strength
and temperature. We find that the breathing mode anomaly depends sensitively
on both interaction strength and temperature. In particular, in the
strongly interacting regime we predict a significant down-shift of
the breathing mode frequency, below the scale invariant value of $2\omega_{0}$
for temperatures of order the Fermi temperature.
\end{abstract}

\maketitle

\section{Introduction}

Fermi gases in two dimensions are of significant importance in understanding
many-body systems \cite{loktev2001,Wen2006}. Ultracold atomic Fermi
gases near a Feshbach resonance offer a new type of strongly interacting
quantum system, where experimentalists have precise control over almost
all the physical properties of the system, such as interaction strength,
particle number, and dimension. Interactions are described through
a zero-range delta-potential, which in two dimensions is known to
be an example of a quantum anomaly \cite{Holstein1993}. The classical
symmetry of a system interacting through a delta potential in two
dimensions is scale invariant under the transformation $\mathbf{r}\rightarrow\lambda\mathbf{r}$,
and an analysis of the scattering properties of a quantum system shows
that it is divergent and must be regularized \cite{Adhikari1986,Levinsen2015,turlapov_rev}.
This well known regularization introduces a new length scale, the
two-dimensional (2D) scattering length $a_{2D}$, and the scale invariance
of the classical theory has been broken. The quantum anomaly can be
seen through the breathing mode of trapped ultracold gases \cite{Rosch97,Olshanii2010}.
Harmonically trapped gases posses a hidden SO(2,1) symmetry \cite{Werner2006},
which excites a breathing mode at a frequency of $\omega=2\omega_{0}$,
where $\omega_{0}$ is the frequency of the trapping potential. The
breaking of scale invariance through the delta-potential interaction
will excite a breathing mode, whose frequency is dependent on the
regularized 2D scattering length, $a_{2D}$. The advent of 2D ultracold
gases realized for fermions \cite{Feld2011,makhalov2014,Fenech2016,Murthy2015,Boettcher2016}
and bosons \cite{Chevy2002,Zaremba2002,Rath2011,Dalibard2011} are
ideal systems for measuring the breathing mode and the anomalous breaking
of scale invariance.

In Fermi gas experiments the breathing mode and anomalous corrections
have been studied by Refs. \cite{Vogt2012,Baur2013}, where the breathing
mode frequency shifts - away from the scale invariance value of $2\omega_{0}$
- were of order a few percent, but the errors within the experiment
made the observation of the anomaly inconclusive. It was argued that
at the temperature of the experiment, $T/T_{F}\simeq0.42$, the anomaly
was damped, and that we would expect at lower temperatures the frequency
shifts of the breathing mode to be more pronounced. Theoretically
the breathing mode of harmonically trapped 2D gases has been extensively
studied at zero temperature \cite{Hofmann2012,Taylor2012,Gao2012}
using quantum Monte Carlo to determine the equation of state \cite{Bertaina2011},
and at finite temperature \cite{Schaefer2D} with the high-temperature
virial expansion up to the second order. In all cases, the breathing
mode frequency can be calculated through a variational approach of
the hydrodynamic equations \cite{Griffin97,Taylor2008,Taylor2009,Hu2014,Rosi2015}.
At $T=0$ the breathing mode shifts from the scale invariant value
found by Refs. \cite{Hofmann2012,Gao2012} are positive for all interaction
strengths at the crossover from a Bose-Einstein condensate (BEC) to
a Bardeen-Cooper Schrieffer (BCS) superfluid, and in the strongly interacting
regime they can reach approximately 10\% of the scale invariant result.
The virial expansion analysis at finite temperature done by Ref. \cite{Schaefer2D}
focused on the comparison to the experiment of Ref. \cite{Vogt2012}
and did not examine the role of temperature on the breathing mode.
To address the temperature effect, it necessarily requires sufficient
knowledge about the equation of state of a strongly interacting 2D
Fermi gas at finite temperature, which unfortunately remains a grand
challenge both experimentally \cite{Fenech2016,Boettcher2016} and
theoretically \cite{Mulkerin2015}.

In Bose gas experiments \cite{Chevy2002,Zaremba2002} the breathing
mode was measured for only weakly interacting systems where the system
appears to be scale invariant and robust to temperature, and no breathing
mode anomaly has been observed. Theoretically, it was found that there
is a temperature dependence of the breathing mode in the weakly interacting
regime \cite{Gies2004}, however it should be noted that this deviation
is not due to the breaking of scale invariance and is a result of
a small deviations from the effective harmonic trapping potential.

In this work, we aim to present a systematic investigation of the
finite-temperature breathing mode of a strongly interacting 2D Fermi
gas, by gathering the most advanced knowledge of the 2D homogeneous
equation of state developed in some recent theoretical works \cite{Liu20102D,Pietil2012,watanabe,Ngamp13,Bauer2014,Marsilgio15,Mulkerin2015,Mulkerin2017superfluid}.
At low temperature, we consider the non-self-consistent pair-fluctuation
theory by Nozières and Schmitt-Rink (NSR) \cite{nozieres1985bose,randeria1989bound}
and the self-consistent $T$-matrix approximation \cite{Haussmann1993}
in the normal state, taking the effects of pairs explicitly into account.
In the high temperature and weakly interacting regimes we employ the
virial expansion \cite{Liu2013} to second and third order \cite{Liu20102D,Ngamp13}.
Using the local density approximation with the homogeneous equation
of state we calculate the collective modes of a trapped system using
a variational approach at a given trap temperature and interaction
strength \cite{Hu2014}. We find theoretically that the breathing
mode anomaly is temperature dependent as well as interaction dependent.
In the weakly interacting BCS regime, the breathing mode is reduced
towards the scale invariant result of $2\omega_{0}$, as temperature
is increased. In the strongly interacting regime and in the high temperature
regime, well above the superfluid temperature $T_{c}$, the breathing
mode lowers below the scale invariant value, indicating the importance
of pair formation at high temperature.

This paper is organized as follows. In Sect.~\ref{Sec:theory} we
introduce a diagrammatic approach to the pressure equation of state
and the virial expansion. In Sect.~\ref{Sec:Hyrdo} we introduce
the variational formalism for ultracold gases and derive the thermodynamic
properties for the calculation of the breathing mode. In Sect.~\ref{Sec:res}
we discuss the results of the breathing mode. We present a brief conclusion
and outlook in Sect.~\ref{Sec:conc}. The two appendices (Appendix
A and Appendix B) are devoted to the details of variationally solving
the hydrodynamic equation.

\section{Theory}

\label{Sec:theory}

In this section, for self-containedness we review the theories of
a strongly interacting Fermi gas in two dimensions \cite{Mulkerin2015}.
At low temperature, we focus on the non-self-consistent NSR approach,
since it provides the easiest way to take into account strong pair-fluctuations
in 2D, with the most stable numerical outputs for the equation of
state. At high temperature, we introduce a Páde approximation to re-organize
the virial expansion series, which may extend the applicability of
virial expansion towards the low-temperature regime. It is worth noting
that for a 2D Fermi gas at finite temperature (i.e., $0<T\lesssim T_{F}$),
all the strong-coupling theories so far are \emph{qualitative} only.
It is notoriously difficult to carry out numerically accurate quantum
Monte Carlo simulations at finite temperature due to the Fermi sign
problem, even in the normal state \cite{anderson2015}. 

\subsection{Many-body $T$-matrix theories}

In order to explore the finite temperature behavior of 2D Fermi gases
at the BEC-BCS crossover, following NSR we consider the contribution
of pairing fluctuations to the thermodynamic potential for a given
temperature $T$ and binding energy $\varepsilon_{B}=\hbar^{2}/(Ma_{2D}^{2})$
through the functional integral formulation, which has been extensively
studied in both two and three dimensions in the normal and superfluid
states \cite{Engelbrecht,Diener2008,Hu2010,Pietil2012,Klimin2012,He2015,Bighin2015,Mulkerin2016,Mulkerin2017superfluid}.
The equation of state is found through the thermodynamic potential
$\Omega=-k_{{\rm B}}T\ln\mathcal{Z}$ where the partition function
is $\mathcal{Z}=\int\mathcal{D}\left[\psi_{\sigma},\bar{\psi}_{\bar{\sigma}}\right]e^{-S\left[\psi_{\sigma},\bar{\psi}_{\bar{\sigma}}\right]}$,
and $\psi_{\sigma}$ and $\bar{\psi}_{\bar{\sigma}}$ are independent
Grassmann fields representing fermionic species for each spin degree
of freedom, $\sigma=\uparrow,\downarrow$, of equal mass $M$. The
partition function is defined through the action
\begin{alignat}{1}
S=\int_{0}^{\hbar\beta}d\tau\left[\int d\mathbf{r}\sum_{\sigma}\bar{\psi}_{\sigma}(x)\partial_{\tau}\psi_{\sigma}(x)+\mathcal{H}\right],
\end{alignat}
and the single channel Hamiltonian given by 
\begin{alignat}{1}
\mathcal{H}=\bar{\psi}_{\sigma}(x)\mathcal{K}\psi_{\sigma}(x)-U_{0}\bar{\psi}_{\uparrow}(x)\bar{\psi}_{\downarrow}(x)\psi_{\downarrow}(x)\psi_{\uparrow}(x),
\end{alignat}
where $\mathcal{K}=-\hbar^{2}\nabla^{2}/(2M)-\mu$, $\beta=(k_{{\rm B}}T)^{-1}$,
$\mu$ is the chemical potential, and $x=(\mathbf{x},\tau)$ for position
$\mathbf{x}$ and imaginary time $\tau$. We take a contact interaction
with $U_{0}>0$, which has known divergences and can be fixed through
renormalizing in terms of the bound state energy via the relation,
\begin{alignat}{1}
\frac{1}{U_{0}}=\sum_{\mathbf{k}}\frac{1}{2\epsilon_{\mathbf{k}}+\varepsilon_{B}},
\end{alignat}
where $\epsilon_{\mathbf{k}}=\hbar^{2}\mathbf{k}^{2}/(2M)$. Using
the Hubbard-Strantonivich transformation to write the action in terms
of the bosonic field and integrating out the fermionic Grassmann fields,
at the Gaussian fluctuation level in the normal state we obtain the
thermodynamic potential, 
\begin{alignat}{1}
\Omega=\Omega_{0}-\sum_{\mathbf{q},\nu_{n}}\ln\left[-\Gamma_{0}^{-1}(\mathbf{q},i\nu_{n})\right].
\end{alignat}
where $\Omega_{0}=2\sum_{\mathbf{k}}\ln(1+e^{-\beta\xi_{\mathbf{k}}})$
is the non-interacting thermodynamic potential and $\xi_{\mathbf{k}}=\epsilon_{\mathbf{k}}-\mu$.
The many-body vertex function, $\Gamma_{0}(\mathbf{q},i\nu_{n})$,
for bosonic Matsubara frequencies $\nu_{n}=2n\pi/\beta$, is given
by 
\begin{alignat}{1}
\Gamma_{0}^{-1}(\mathbf{q},i\nu_{n})=\sum_{\mathbf{k}}\left[\frac{1-f(\xi_{\mathbf{k}})-f(\xi_{\mathbf{k}+\mathbf{q}})}{i\nu_{n}-\xi_{\mathbf{k}+\mathbf{q}}-\xi_{\mathbf{k}}}+\frac{1}{2\epsilon_{\mathbf{k}}+\varepsilon_{B}}\right]
\end{alignat}
where we have defined the Fermi distribution $f(x)=1/(e^{\beta x}+1)$.
The system can be viewed as a non-interacting mixture of fermions
and pairs. We can analytically continue the Matsubara summation of
the bosonic frequencies and find the contribution to the pairing fluctuations,
writing the thermodynamic potential as $\Omega=\Omega_{0}+\Omega_{{\rm GF}}$,
\begin{alignat}{1}
\Omega_{{\rm GF}}=-\sum_{\mathbf{q}}\intop_{-\infty}^{+\infty}\frac{d\omega}{\pi}b(\omega){\rm Im}\ln\left[-\Gamma_{0}^{-1}(\mathbf{q},\omega+i0^{+})\right],
\end{alignat}
where $b(\omega)=1/(e^{\beta\omega}-1)$. The pressure equation of
state is found from the thermodynamic potential for a given temperature
$T$ and binding energy $\varepsilon_{B}$ by $\Omega=-PV$, where
$V$ is the volume of the system, and the density equation of state
is found by satisfying $n=-\partial\Omega/\partial\mu$. The dimensionless
pressure equation of state is defined, 
\begin{alignat}{1}
\frac{P\lambda_{T}^{2}}{k_{{\rm B}}T} & =f_{p}\left(\frac{\mu}{k_{{\rm B}}T}\right)
\end{alignat}
where the thermal wavelength is $\lambda_{T}=\sqrt{2\pi\hbar^{2}/(Mk_{B}T)}$,
and pressure equation of state is related to the density through $f_{n}=\partial f_{p}/\partial(\beta\mu)$.
This derivation of the thermodynamic potential and calculation of
the density equation of state is equivalent to taking the $T$-matrix
approximation with bare fermionic Green functions in the truncated
Schwinger-Dyson equations \cite{serene1989}. For this reason, the
above NSR approach is alternatively termed as the non-self-consistent
$T$-matrix approximation. 

The NSR equation of state can be considered against other $T$-matrix
schemes such as the Luttinger-Ward theory, which is a fully self-consistent
calculation of the many-body Green's function and we will refer to
as $GG$ theory \cite{Haussmann93,Bauer2014,Mulkerin2015}. The NSR
method in two-dimensions has been found to underestimate the density
when compared to current experimental results and $GG$ theory \cite{Mulkerin2015},
for the calculation of the hydrodynamic equations the NSR method is
advantageous over other $T$-matrix schemes as it allows for the calculation
of the thermodynamic properties and collective modes at a fixed $\beta\mu$
and $\beta\varepsilon_{B}$.

\subsection{Virial expansion}

In the high temperature and weakly interacting limits we calculate
the equation of state through the virial expansion, where the thermodynamic
potential is expanded in powers of the fugacity, $z=e^{\beta\mu}$
\cite{Liu2013}. This allows for an exact calculation of the thermodynamic
properties and the breathing mode. It is straight forward to calculate
the high temperature regime through the virial expansion, here we
present some details of the calculation of the virial coefficients
and densities and further reading can be found in Refs.~\cite{Liu2013,Barth2014,Ngamp13,Leyronas2011}.

Let us consider the virial expansion up to third order, the pressure
and density equations of state are 
\begin{alignat}{1}
f_{p} & =\int_{0}^{\infty}dt\ln\left[1+ze^{-t}\right]+\Delta b_{2}z^{2}+\Delta b_{3}z^{3}+\dots,\nonumber \\
f_{n} & =\frac{\partial f_{p}}{\partial\beta\mu}=\ln\left[1+z\right]+2\Delta b_{2}z^{2}+3\Delta b_{3}z^{3}+\dots,
\end{alignat}
where $\Delta b_{2}$ and $\Delta b_{3}$ are the second and third
virial coefficients respectively. It is important to note that the
virial coefficients are functions of the dimensionless binding energy,
$\beta\varepsilon_{B}$, only. We calculate the second order virial
coefficient through the relation \cite{Ngamp13}, 
\begin{alignat}{1}
\Delta b_{2}=e^{\beta\varepsilon_{B}}-\int_{0}^{\infty}\frac{dp}{p}\frac{e^{p^{2}}}{\pi^{2}+\ln\left[p^{2}/(\beta\varepsilon_{B})\right]},\label{eq:b2}
\end{alignat}
and the third order coefficient, $\Delta b_{3}$, has been tabulated
as a function of $\beta\varepsilon_{B}$ in the range $\beta\varepsilon_{B}=(0.001,10.0)$
\cite{Fenech2016}. The third order expansion of the pressure and
density equations of state are divergent when $z>1$, however through
a Páde expansion we may overcome this divergence and expand the regime
of the equation of state \cite{Hupade}. Specifically the third order
reduced pressure through the Páde expansion is, 
\begin{alignat}{1}
\frac{P}{P_{0}}=\frac{1+\left(b_{2}^{(1)}+\Delta b_{2}-\Delta b_{3}/\Delta b_{2}\right)e^{\beta\mu}+\cdots}{1+\left(b_{2}^{(1)}-\Delta b_{3}/\Delta b_{2}\right)e^{\beta\mu}+\cdots}
\end{alignat}
where $b_{2}^{(1)}=-1/4$ is the second virial coefficient of an ideal
2D Fermi gas and $P_{0}=-2\pi\lambda_{T}^{-4}{\rm Li}_{2}(-e^{\beta\mu})$
is the ideal pressure with ${\rm Li}_{2}$($x$) being the polylogarithm.
Although the expansion can be found for any value of the fugacity,
$z$, and the pressure equation of state no longer diverges, there
is no \textit{a}\emph{ priori} reason for the expansion to be physically
correct. We note that the determination of higher order terms such
as $e^{2\beta\mu}$ requires knowledge of the fourth and fifth virial
coefficients.

The virial expansion is valid in the BEC limit where $\varepsilon_{B}$
becomes large and the chemical potential approaches $\mu\rightarrow-\varepsilon_{B}/2$.
In this limit the virial expansion would appear to be valid in the
limit $T\rightarrow0$, however the dimer contribution to the virial
coefficients dominates and the expansion of the thermodynamic potential
in this regime should be in terms of a shifted fugacity, $z^{({\rm Bose)}}=e^{\beta\varepsilon_{B}/2}z$,
the criterion for the validity of the expansion then becomes $\mu<-\varepsilon_{B}/2$.
In this expansion the virial coefficients correspond to $b_{j}^{({\rm Bose)}}=e^{-j\beta\varepsilon_{B}/2}b_{j}$.
An expansion of the Bose thermodynamic potential, taking the effective
dimer-dimer scattering length $a_{dd}\simeq0.56a_{2D}$ \cite{He2015,Salsnich2015},
is also dominated by the large binding energy and has the same behavior
in the BEC regime.

\begin{figure}
\centering{}\includegraphics[width=0.95\columnwidth]{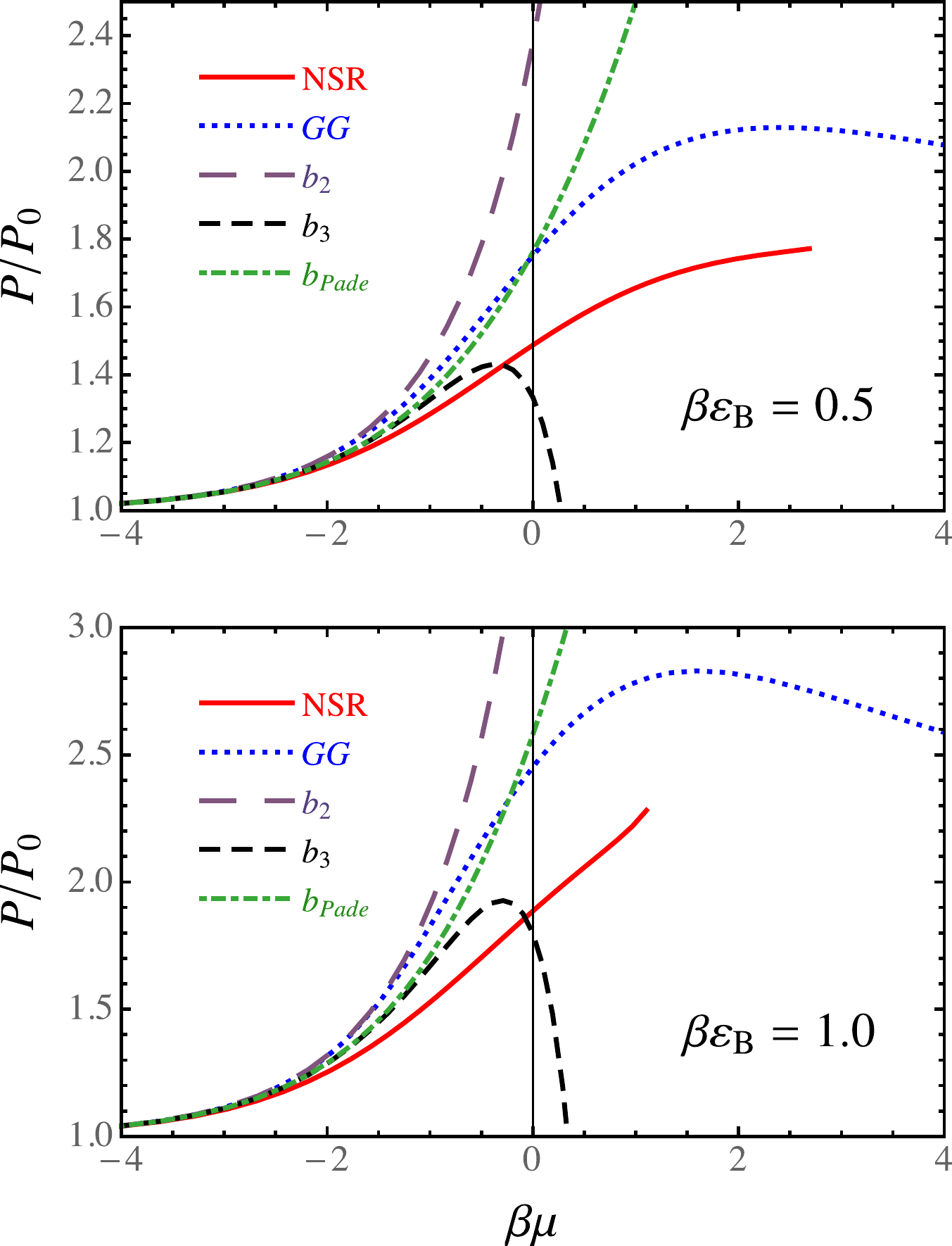}
\caption{(color online). The pressure equation of state, in units of the pressure
of an ideal Fermi gas $P_{0}$, as a function of the chemical potential
at interaction strengths $\beta\varepsilon_{B}=0.5$ and $\beta\varepsilon_{B}=1.0$
for the NSR (red solid) and $GG$ (blue dotted) $T$-matrix theories,
second (purple long-dashed) and third (black dashed) order virial
expansions, and the Páde expansion of the third order virial (green
dot-dashed). \label{fig:peos}}
\end{figure}

\subsection{The pressure equation of state}

The pressure equation of state plays a key role in determining the
collective modes of a strongly interacting Fermi gas in the hydrodynamic
regime \cite{Taylor2009}. To illustrate the structure of the equation
of state before we calculate the collective modes we plot the reduced
pressure equation of state, $P/P_{0}$, in Fig.~\ref{fig:peos}.
We show the pressure equation of state for the NSR (red solid) and
$GG$ (blue dotted) $T$-matrix theories, the second (purple long-dashed)
and third (black short dashed) virial expansions, and the P\'ade expansion
(green dot dashed) of the third order virial for interaction strengths
of $\beta\varepsilon_{B}=0.5$ and $\beta\varepsilon_{B}=1.0$, as
a function of the reduced chemical potential, $\beta\mu$. As has
been discussed in Refs.~\cite{Bauer2014,Mulkerin2016} the equation
of state exhibits a non-trivial maximum as the temperature reduces
($\beta\mu\rightarrow\infty$), and in Fig.~\ref{fig:peos} we see
that the pressure equations of state for the NSR and $GG$ theories
begin to reduce at low temperatures. The pressure is underestimated
by the NSR when compared to the $GG$ and virial expansions. The virial
expansions do not have the non-trivial dependence on the temperature,
and are incorrect in the low temperature regime ($\beta\mu\rightarrow\infty$)
where they overestimate the pressure.

We see in Fig.~\ref{fig:peos} that the NSR theory breaks down at
higher temperatures compared to the $GG$ theory, making the range
of temperatures in the calculation of the breathing mode smaller.
It is well known that two-dimensional $T$-matrix theories cannot
predict a transition to superfluidity due to the breakdown of long-range
order at any finite temperature, however the NSR theory still suffers
from a loss of accuracy as the temperature is reduced and the chemical
potential approaches half the binding energy (for a detailed analysis
see Ref.~\cite{loktev2001,Marsilgio15}). The $GG$ theory is more
robust and the equation of state can be calculated to lower temperatures,
however we focus on calculating the collective modes of the two-dimensional
gas through the NSR theory as we can calculate the numerical derivatives
with respect to $\beta\varepsilon_{B}$ and $\beta\mu$. The equation
of state for the $GG$ theory is found for a fixed $\beta\varepsilon_{B}$
and we iteratively find a chemical potential that satisfies the corresponding
number equation, making the calculation of thermodynamic properties
heavily reliant on numerical interpolation. We expect from previous
studies that the breathing mode will not be too sensitive to the equation
of state \cite{Heiselberg2004,Hui2004}. Therefore, the use of the
equations of state from different theories will not greatly affect
the calculation of the breathing mode frequency and the \emph{qualitative}
behavior of the breathing mode could be captured.

\section{Hydrodynamic Framework}

\label{Sec:Hyrdo}

We follow the work of \cite{Taylor2008,Taylor2009} to calculate the
finite temperature collective modes of a two-dimensional Fermi gas
trapped in a harmonic potential $V_{{\rm tr}}=\frac{1}{2}M\omega_{0}^{2}r^{2}$.
From the equation of state of the homogeneous 2D Fermi gas and the
local density approximation, the collective modes of a Fermi gas can
be found for frequency $\omega$ and temperature $T$ by minimizing
a variational action, which in terms of the normal state displacement
fields $\mathbf{u}_{n}(\mathbf{r})$ takes the form \cite{Hu2014},
\begin{alignat}{1}
S^{(2)}= & \frac{1}{2}\int d\mathbf{r}\biggl[\omega^{2}\rho_{0}\mathbf{u}_{n}(\mathbf{r})^{2}-\frac{1}{\rho_{0}}\left(\frac{\partial P}{\partial\rho}\right)_{\bar{s}}\left(\delta\rho\right)^{2}\nonumber \\
 & -2\rho_{0}\left(\frac{\partial T}{\partial\rho}\right)_{\bar{s}}\delta\rho\delta\bar{s}-\rho_{0}\left(\frac{\partial T}{\partial\bar{s}}\right)_{\bar{s}}\left(\delta\bar{s}\right)^{2}\biggl].\label{eq:HydroAction}
\end{alignat}
We define the total mass density at equilibrium, $\rho(\mathbf{r})\equiv Mn(\mathbf{r})=\rho_{0}$,
the local pressure $P(\mathbf{r})=P_{0}$, the entropy per unit mass
$\bar{s}(\mathbf{r})\equiv s/\rho=\bar{s}_{0}$, and $\delta\rho(\mathbf{r})=-\nabla\cdot\left(\rho_{0}\mathbf{u}_{n}\right)$
and $\delta\bar{s}(\mathbf{r})=-\mathbf{u}_{n}\cdot\nabla\bar{s}_{0}$
are the density and entropy fluctuations, respectively. Taking the
variation of the action with respect to $\mathbf{u}_{n}$ we arrive
at 
\begin{alignat}{1}
 & \omega^{2}\rho_{0}\mathbf{u}_{n}+\nabla\left[\rho_{0}\left(\frac{\partial P}{\partial\rho}\right)_{\bar{s}}\left(\nabla\cdot\mathbf{u}_{n}\right)\right]+\nonumber \\
 & \nabla\cdot\left(\rho_{0}\mathbf{u}_{n}\right)\frac{\nabla V_{{\rm tr}}}{M}+\nabla\left[\nabla P_{0}\cdot\mathbf{u}_{n}\right]=0,\label{eq:Euler}
\end{alignat}
which is Euler's equation as first derived in Re.~\cite{Griffin97}.
For an untrapped gas the solution of Eq.~\eqref{eq:Euler} is a wave
vector $q$, with dispersion $\omega=c_{n}q$ and $c_{n}$ is the
speed of sound, 
\begin{alignat}{1}
c_{n}=\sqrt{\frac{1}{m}\left(\frac{\partial P}{\partial n}\right)_{\bar{s}}},
\end{alignat}
where 
\begin{alignat}{1}
n\left(\frac{\partial P}{\partial n}\right)_{\bar{s}} & =n\left(\frac{\partial P}{\partial n}\right)_{s}+s\left(\frac{\partial P}{\partial s}\right)_{n},\nonumber \\
 & =n\frac{\partial(P,s)/\partial(\mu,T)}{\partial(n,s)/\partial(\mu,T)}+s\frac{\partial(P,n)/\partial(\mu,T)}{\partial(s,n)/\partial(\mu,T)},\nonumber \\
 & =\frac{n(P_{\mu}s_{T}-P_{T}s_{\mu})-s(P_{\mu}n_{T}-P_{T}n_{\mu})}{(n_{\mu}s_{T}-n_{T}s_{\mu})}.
\end{alignat}
For convenience we use the shorthand notation $P_{\mu}\equiv(\partial P/\partial\mu)_{T}$,
$s_{T}\equiv(\partial s/\partial T)_{\mu}$, etc, and in dimensionless
form we obtain 
\begin{alignat}{1}
\left(\frac{\partial P}{\partial n}\right)_{\bar{s}} & =k_{{\rm B}}T\frac{(\tilde{P}_{\mu}\tilde{s}_{T}-\tilde{P}_{T}\tilde{s}_{\mu})-\tilde{\bar{s}}(\tilde{P}_{\mu}\tilde{n}_{T}-\tilde{P}_{T}\tilde{n}_{\mu})}{(\tilde{n}_{\mu}\tilde{s}_{T}-\tilde{n}_{T}\tilde{s}_{\mu})}.
\end{alignat}
All of the thermodynamic quantities can then be found from the pressure
equation of state by taking partial derivatives with respect to the
reduced chemical potential, $\beta\mu$ and interaction strength,
$\beta\varepsilon_{B}$.

We calculate the thermodynamic potential of a harmonically trapped
gas through the local density approximation, $\mu(r)=\mu-V_{{\rm tr}}(r)$,
and write the local pressure and number density using the universal
functions, 
\begin{alignat}{1}
P(r)=\frac{k_{{\rm B}}T}{\lambda_{{\rm T}}^{2}}f_{p}\left[\frac{\mu(r)}{k_{{\rm B}}T}\right],\,\,n(r)=\frac{1}{\lambda_{{\rm T}}^{2}}f_{n}\left[\frac{\mu(r)}{k_{{\rm B}}T}\right],
\end{alignat}
where the local chemical potential is $\mu(r)$ and the Fermi temperature
is defined as, 
\begin{alignat}{1}
k_{{\rm B}}T_{{\rm F}}=\hbar(2N\omega_{0}^{2})^{1/2}.
\end{alignat}
Using the number equation $N=\int d\mathbf{r}n(r)$ we relate the
reduced chemical potential $\beta\mu$ to the reduced temperature
in the trap $T/T_{F}$ by, 
\begin{alignat}{1}
\frac{T}{T_{F}}=\left[2\int_{0}^{\infty}dyf_{n}\left(\beta\mu-y\right)\right]^{-1/2}.\label{eq:temp}
\end{alignat}
For a given trap temperature we need the dimensionless chemical potential,
$\beta\mu_{c}$, which satisfies Eq.~\eqref{eq:temp} (as temperature
reduces $\beta\mu_{c}\rightarrow\infty$). To minimize the action
in Eq.~\eqref{eq:HydroAction} we expand the displacement field in
a variational (polynomial) basis by using the following ansatz, 
\begin{alignat}{1}
\mathbf{u}_{n}(\mathbf{r})=\sum_{n=0}^{N_{\perp}}A_{n}r^{n+1},
\end{alignat}
where we can increase the precision of the variational calculation
by increasing the number of basis functions, $N_{\perp}$. Expanding
the action we get, 
\begin{alignat}{1}
S_{n}^{(2)}=\frac{1}{2}\left[A_{0}^{*},\cdots,A_{N_{\perp}}^{*}\right]\mathbf{A}(\omega)\left[A_{0},\cdots,A_{N_{\perp}}\right]^{{\rm T}},
\end{alignat}
where $\mathbf{A}(\omega)$ is a $N_{\perp}\times N_{\perp}$ matrix,
\begin{alignat}{1}
A_{nm}(\omega)=\omega^{2}M_{nm}-K_{nm}.
\end{alignat}
Here we have defined the weighted mass moments, $\mathbf{M}=M_{nm}$,
and spring constants, $\mathbf{K}=K_{nm}$ given in Appendix \ref{app:var}.
To find the mode frequencies we need to solve the matrix equation,
\begin{alignat}{1}
\mathbf{A}(\tilde{\omega})\mathbf{x}=0,\label{eq:maindet}
\end{alignat}
where the vector of displacement fields are given by, $\mathbf{x}=\left[A_{0},\dots,A_{n},\dots\right]^{{\rm T}}$,
and we write the matrix $\mathbf{A}(\tilde{\omega})=\mathbf{M}\tilde{\omega}^{2}-\mathbf{K}$
in terms of the dimensionless frequency $\tilde{\omega}=\omega/\omega_{0}$.
The details of this calculation are given in Appendix \ref{App:solvedet}.
The breathing mode is the lowest frequency found from solving Eq.~\eqref{eq:maindet}.
The variational result converges quickly, where the $n=0$ giving
the most significant contribution.

\begin{figure}
\centering{}\includegraphics[width=1\columnwidth]{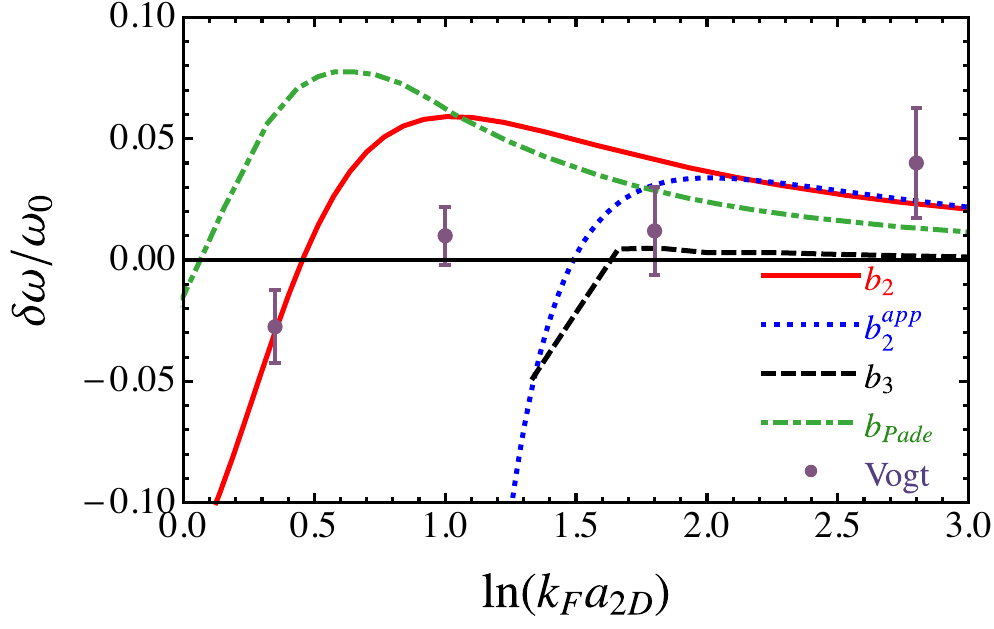}
\caption{(color online). The frequency shift of the breathing mode from the
scale invariant value, $\delta\omega=\omega-2\omega_{0}$, at temperature
$T/T_{{\rm F}}=0.42$ as a function of interaction strength $\ln\left(k_{{\rm F}}a_{2D}\right)$
for the second order virial (red solid), dilute limit second order
virial (blue dotted) \cite{Schaefer2D}, third order virial (black
dashed), Páde virial expansion (green dot-dashed), and experimental
results from Ref.~\cite{Vogt2012} (symbols).}
\label{fig:breath42vir} 
\end{figure}

\section{Results and Discussion}

\label{Sec:res} 

We now turn to examining the breathing mode at finite temperature.
We first consider the results obtained by using the virial expansion
approach, contrasted to the experimental data of Ref.~\cite{Vogt2012}
and the previous virial results of Ref.~\cite{Schaefer2D}, and then
compare the predictions from the different $T$-matrix theories. We
finally check the breathing mode in the high temperature regime, where
the virial expansion is more reliable and we examine the role of pairing
at high temperature.

\subsection{Virial comparison}

In Fig.~\ref{fig:breath42vir} we compare the frequency shift of
the breathing mode from the scale invariant value, $\delta\omega=\omega-2\omega_{0}$,
as a function of interaction strength $\ln\left(k_{{\rm F}}a_{2D}\right)$
on the BCS side for the virial expansion at a temperature of $T/T_{F}=0.42$.
The use of virial expansion at this (low) temperature could be questionable.
However, we present the comparison for the purpose of making connect
to the experimental measurement \cite{Vogt2012} and also making contact
with the previous virial expansion study \cite{Schaefer2D}. We compute
the breathing mode using the second order viral (red solid), third
order virial (black dashed), Páde expansion of the third order virial
(green-dot dashed), a dilute expansion of the second order virial
as found in Ref.~\cite{Schaefer2D} (blue dotted), and the experimental
results of Ref.~\cite{Vogt2012} (symbols). 

The breathing mode found in Ref~\cite{Schaefer2D} is calculated
from a second order expansion of the equation of state, however the
authors \emph{further} approximate the trap density and the speed
of sound in the dilute limit, computing the breathing mode to be 
\begin{alignat}{1}
\frac{\omega^{2}}{4\omega_{0}^{2}}=1-\frac{1}{8}\frac{T_{{\rm F}}^{2}\varepsilon_{B}^{2}}{T^{4}}\frac{\partial^{2}\Delta b_{2}}{\partial(\beta\varepsilon_{B})^{2}},
\end{alignat}
where $\Delta b_{2}(\beta\varepsilon_{B})$ is given by Eq.~\eqref{eq:b2}.
We see that our second order virial calculation extends to stronger
interaction strengths before lowering, however we emphasize that at
this temperature the virial expansion is not accurate and the results
should be treated as qualitative only. We see that at the reduced
trap temperature of $T/T_{{\rm F}}=0.42$ over the interaction regime
the virial expansions of second order, third order, and Páde expansion
are not the same. This can be understood by looking at the pressure
equation of state in Fig.~\ref{fig:peos}, the critical chemical
potential $\beta\mu_{c}$ needed to determine the trap temperature
in Eq.~\eqref{eq:temp} is in the regime where the second and third
expansions differ significantly, and where the third order is diverging.
Although we can calculate the thermodynamic properties for the speed
of sound within the virial expansion, for this temperature and interaction
regime, the breathing mode results are only qualitative.

\subsection{$T$-matrix results at low temperature}

\begin{figure}[t]
\centering{}\includegraphics[width=1\columnwidth]{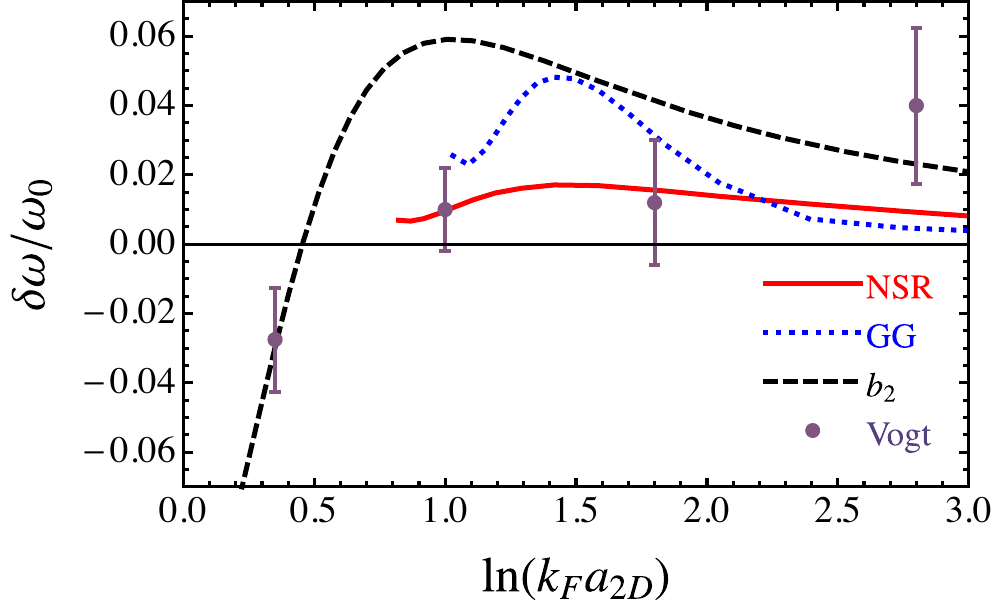}
\caption{(color online). The frequency shift of the breathing mode from the
scale invariant value, $\delta\omega=\omega-2\omega_{0}$, at temperature
$T/T_{{\rm F}}=0.42$ as a function of interaction strength $\ln\left(k_{{\rm F}}a_{2D}\right)$
for the NSR (red solid) and $GG$ (blue dotted) $T$-matrix theories,
second order virial (black dashed), and experimental results from
Ref.~\cite{Vogt2012} (symbols).}
\label{fig:breath42} 
\end{figure}

The $T$-matrix theories take
into account the many-body effects and pairing fluctuations, extending the 
equation of state found through the virial expansions to lower temperatures. In Fig.~\ref{fig:breath42}
we compare the frequency shifts of the breathing mode $\delta\omega=\omega-2\omega_{0}$
at a reduced temperature of $T/T_{F}=0.42$ obtained by the NSR (red
solid) and self-consistent $GG$ $T$-matrix (blue dotted) theories,
second order virial expansion (black dashed), and also compare them
to the experimental work of Ref.~\cite{Vogt2012} (symbols). For
each of the frequency shifts there is a maximum, and as the interaction
becomes stronger the frequency shift reduces towards $\delta\omega=0$,
as is seen in the experimental results of Ref.~\cite{Vogt2012}.
The difference between the NSR and $GG$ theories could be due to
the fact that the NSR approach underestimates the pressure. Although
the second order virial expansion is not reliable in this regime,
the qualitative behavior is similar to the NSR and $GG$ theories.

We see in Fig.~\ref{fig:breath42} that the NSR breathing mode breaks
down for an interaction of $\ln\left(k_{{\rm F}}a_{2D}\right)\simeq0.9$
where the critical chemical potential needed for the reduced trap
temperature is too large within the $T$-matrix theory. The $GG$
frequency shift breaks down at $\ln\left(k_{{\rm F}}a_{2D}\right)\simeq1.0$,
and is different to the breakdown of the NSR theory.
For stronger interactions the numerical noise in the calculation of
thermodynamic properties is too large to accurately determine the
speed of sound.

\begin{figure}
\centering{}\includegraphics[width=1\columnwidth]{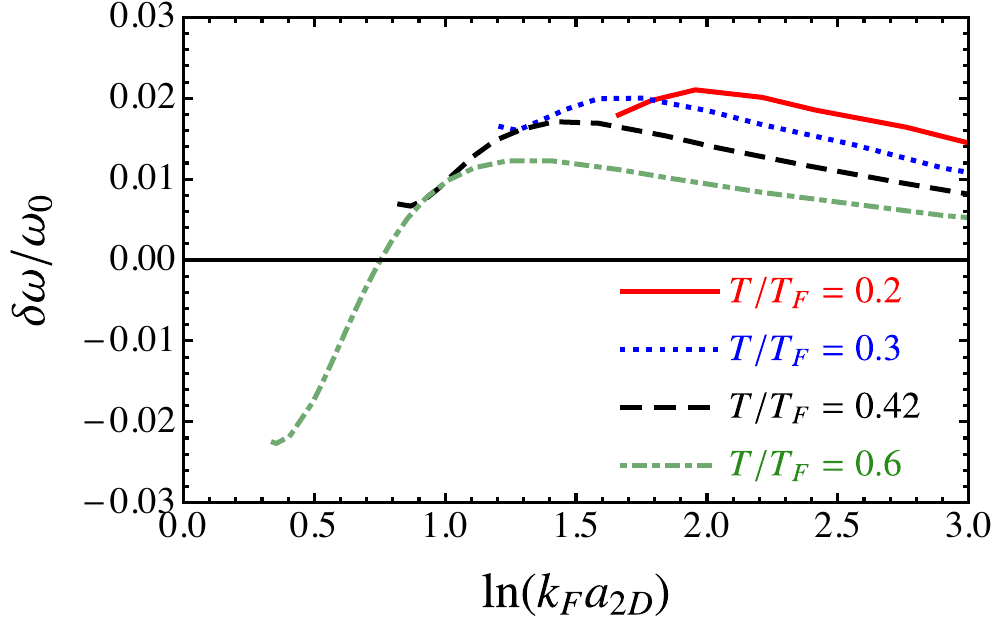}
\caption{(color online). The frequency shift of the breathing mode, $\delta\omega=\omega-2\omega_{0}$,
as a function of interaction strength $\ln\left(k_{{\rm F}}a_{2D}\right)$
for the NSR $T$-matrix at temperature $T/T_{{\rm F}}=0.2$ (red solid),
$T/T_{{\rm F}}=0.3$ (blue dotted), $T/T_{{\rm F}}=0.42$ (black dashed),
and $T/T_{{\rm F}}=0.6$ (green dot-dashed).}
\label{Fig:temps} 
\end{figure}

In Fig.~\ref{Fig:temps} we consider how the frequency shift of the
breathing mode behaves as a function of temperature using the NSR
theory, specifically for $T/T_{F}=0.2$ (red solid), $T/T_{F}=0.3$
(blue dotted), $T/T_{F}=0.42$ (black dashed), and $T/T_{F}=0.6$
(green dot-dashed). As temperature decreases the frequency shift increases,
indicating that as temperature is reduced the breathing mode anomaly
will be larger.
We find that the range of validity of the NSR calculation reduces
with decreasing temperature and the NSR theory also breaks down at
higher temperatures as we increase the binding energy. The frequency
shift at all temperatures considered has a maximum value that lowers
as interaction strength increases, a result consistent in all of the
theories and temperatures considered in this work. Therefore, we believe
that this will be qualitatively true for further experimental checks
at finite temperature. However, for temperatures below $T/T_{F}=0.2$
it is not clear if the frequency shift will lower or be positive over
the whole BEC-BCS crossover.

\begin{figure}
\centering{}\includegraphics[width=1\columnwidth]{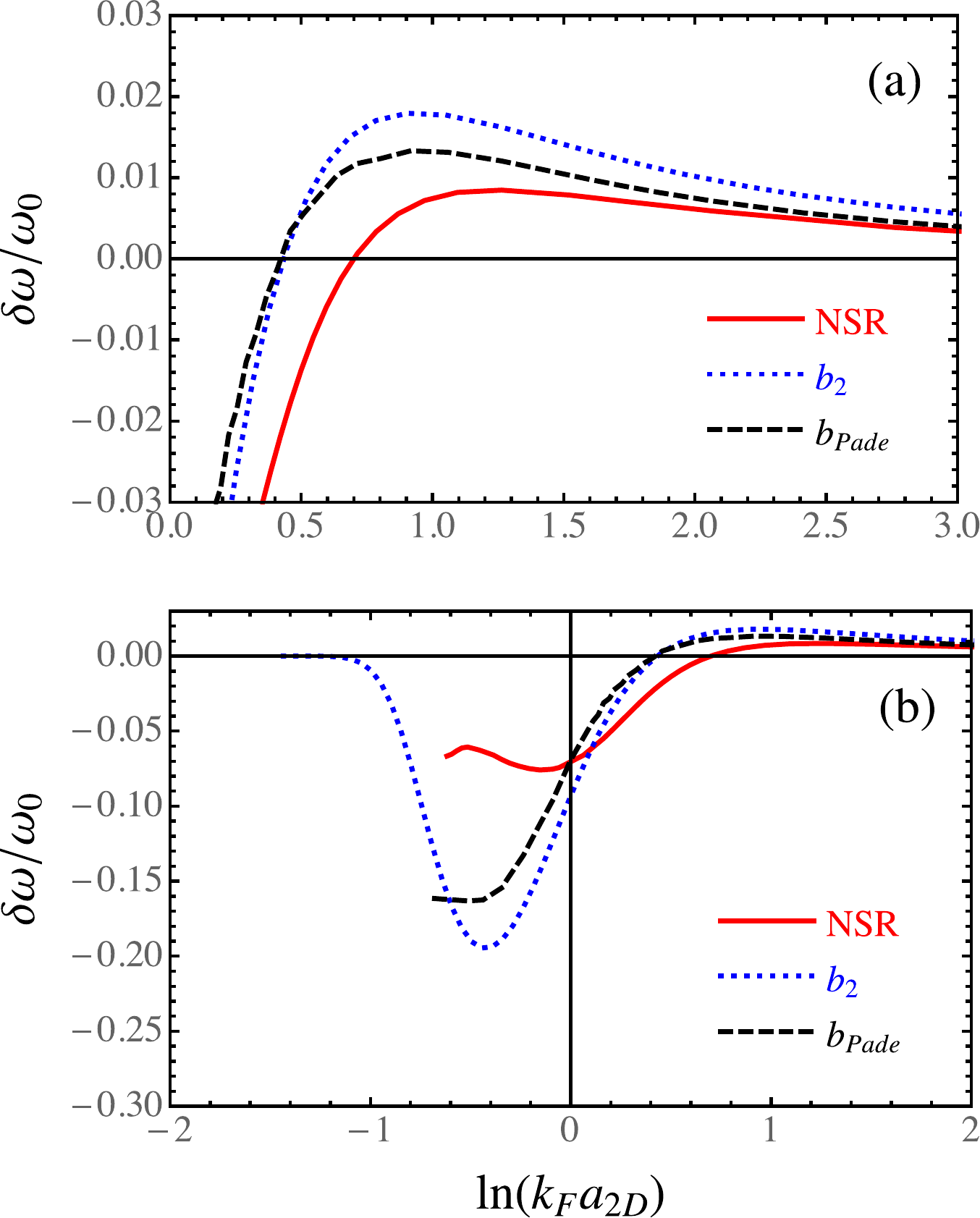}
\caption{(color online). The frequency shift of the breathing mode, $\delta\omega=\omega-2\omega_{0}$,
at temperature $T/T_{{\rm F}}=0.8$ as a function of interaction strength
$\ln\left(k_{{\rm F}}a_{2D}\right)$ for the NSR $T$-matrix (red
solid), second order virial expansion (blue dotted), and third order
Páde virial expansion (black dashed), for (a) the BCS regime and (b)
the BEC-BCS crossover.}
\label{fig:breath08} 
\end{figure}

\subsection{High temperature limit}

For the reduced trap temperature of $T/T_{F}=0.6$ in Fig. \ref{Fig:temps}
the interaction range of validity is increased and we see that the
frequency shift becomes negative before the NSR results break down.
This increased range of validity allows us to examine the breathing
mode and the effects of pairing at temperatures far above the critical
temperature. In two dimensions the role of pairing in the high temperature
regime is a widely discussed area of research \cite{Murthy2017,Feld2011}.

The NSR and virial expansions are valid for a wide range of interactions
at high temperatures and we can cover the BEC-BCS crossover, where
we also \emph{assume} that the hydrodynamic equations are still valid
\cite{Wright2007}. In Fig.~\ref{fig:breath08} we calculate the
breathing mode shift for a temperature of $T/T_{{\rm F}}=0.8$, with
the NSR theory (red solid), second order virial (blue dotted), and
Páde expansion (black dashed). Figure \ref{fig:breath08}(a) shows
that in the BCS limit the frequency shift is approaching the scale
invariant result of $\delta\omega=0$. As we approach the strongly
interacting regime, the frequency shift has a maximum value and then
tends negative, and the breathing mode anomaly seems to be weakened
at first glance. However, expanding the interaction range and looking
at Fig.~\ref{fig:breath08}(b) we see that the NSR, virial, and Páde
expansion predict a significant negative shift in the strongly interacting
regime. The NSR theory finds a breathing mode shift of $~4\%$ and
the second order virial expansion $~10\%$. Going deeper to the BEC
side the NSR and Páde theories break down for $\ln(k_{{\rm F}}a_{2D})\simeq-0.6$,
and the second order virial expansion finds that the breathing mode
approaches the scale invariant value of $2\omega_{0}$ by $\ln(k_{{\rm F}}a_{2D})\simeq-1$. 

We would like to argue that the significant down-shift of the breathing
mode in the strongly interacting regime is due to pairing effects.
On the BEC side of the strongly interacting regime (i.e., $-1<\ln(k_{{\rm F}}a_{2D})<0$
with $\beta\mu<0$), the two-body pairing is captured by the second
order virial expansion or the Páde expansion. Here, it is known that
pairing above $T_{c}$ is without a Fermi surface \cite{Levinsen2015},
and therefore is not associated with the pseudogap. In contrast, on
the BCS side of the strongly interacting regime, the NSR theory predicts
a chemical potential of $\beta\mu\simeq0$, and pair formation is
a many-body effect. Any possible experimental observations of the
predicted down-shift of the breathing mode frequency are therefore
of important for understanding the role of pairing in two-dimensional
Fermi systems.

\section{Conclusion}

\label{Sec:conc}

In summary we have investigated the behavior of the breathing mode
at finite temperature for a strongly interacting 2D Fermi gas at the
BEC-BCS crossover. Using the equation of state found from the NSR
and self-consistent $T$-matrix theories as well as the virial expansion
at different orders for a homogenous Fermi gas, we have predicted
the breathing mode at finite temperature of a trapped gas through
the local density approximation and a variational approach to the
hydrodynamic Euler equation. The use of different theories for a strongly
interacting Fermi gas enable us to paint a broad and qualitative picture
of the breathing mode at finite temperature. Both $T$-matrix theories
and virial expansions consistently show the sensitivity of the quantum
anomaly on the temperature and interaction strength, that is, the
frequency shift of the breathing mode is sensitively dependent on
temperature and interaction. 

On the BCS side, we have predicted that the breathing mode frequency
reduces towards the scale invariant value of $2\omega_{0}$ as temperature
increases and, the quantum anomaly is more prominent at low temperatures,
as one may anticipate. At the typical interaction strength $\ln(k_{{\rm F}}a_{2D})\simeq2$,
the frequency shift predicted by the NSR approach is at the level
of 1\% for temperature up to $0.6T_{F}$. Considering the high experimental
resolution for frequency measurements with cold-atoms, which is about
0.1\% \cite{Tey2013}, this shift is significant enough to
be resolved in future experiments. 

In the strongly interacting regime, we have confirmed that a significant
negative frequency shift at high temperatures, peaking near $\ln\left(k_{{\rm F}}a_{2D}\right)\simeq-0.4$,
as predicted by both NSR and virial expansion theories. This may be
due to the strong pairing effects included in the calculations of
the homogeneous equation of state. We note that, a down-shift of the
breathing mode frequency, below the scale invariant value, was also
predicted by Chafin and Schäfer using the second-order virial expansion
at $T=0.42T_{F}$ \cite{Schaefer2D}. Our more systematic virial expansion
studies, beyond the second order, unambiguously confirm their finding
and establish the qualitative behavior of the breathing mode frequency
at high temperature at the whole BEC-BCS crossover.
\begin{acknowledgments}
We would like to thank T. Peppler, P. Dyke, and C. Vale for their
useful discussions. This research was supported under Australian Research
Council's Discovery Projects funding scheme (project numbers DP140100637,
DP140103231, and DP170104008) and Future Fellowships funding scheme
(project numbers FT130100815 and FT140100003).
\end{acknowledgments}

\appendix

\section{Variational approach}

\label{app:var}

Here we consider in more detail the expressions for the weighted mass
moments, $M_{nm}$, and spring constants, $K_{nm}$. We present a
detailed derivation of the weighted mass moments as this will be instructive
to the choice of units. The weighted mass moments, $M_{nm}$, arise
from the following action term 
\begin{alignat}{1}
\omega^{2}\int d\mathbf{r}\rho_{0}\mathbf{u}_{n}^{2}=A_{n}A_{m}\,\omega^{2}\int d\mathbf{r}\rho_{0}(r)\tilde{r}^{n+m+2},
\end{alignat}
where we have denoted $\tilde{r}\equiv r/R_{{\rm TF}}$ and $R_{{\rm TF}}^{2}=2k_{{\rm B}}T_{{\rm F}}/(m\omega_{0}^{2})$
is the Thomas-Fermi temperature for a zero-temperature noninteracting
Fermi gas. Recalling that within the local density approximation,
we have, $\rho_{0}(r)=Mn_{0}(r)=M\lambda_{T}^{-2}f_{n}\left[(\mu-V_{{\rm tr}}(r))/k_{B}T\right]$,
thus 
\begin{alignat}{1}
\int d\mathbf{r}\rho_{0}(r)\tilde{r}^{k} & =\frac{R_{{\rm F}}^{2}M}{\lambda_{T}^{2}}\int d\tilde{\mathbf{r}}f_{n}\left(\beta\mu-\frac{M\omega_{0}^{2}r^{2}}{2k_{{\rm B}}T}\right)\tilde{r}^{k},\nonumber \\
 & =\frac{R_{{\rm F}}^{2}M2\pi}{\lambda_{T}^{2}}\int d\tilde{r}f_{n}\left(\beta\mu-\frac{\tilde{r}^{2}}{T/T_{F}}\right)\tilde{r}^{k+1}.
\end{alignat}
There is a constant here that will set to 1 as it appears in all of
the equations. As a result of the above calculation we can tie the
reduced temperature $T/T_{F}$ of the system to a given $\mu/k_{{\rm B}}T$
by using the number equation for $N$ atoms, $MN=\int d\mathbf{r}\rho_{0}(r)$,
that is 
\begin{alignat}{1}
N\lambda_{T}^{2} & =\int d\mathbf{r}f_{n}\left(\beta\mu-\frac{M\omega^{2}r^{2}}{2k_{{\rm B}}T}\right),\nonumber \\
\frac{N\lambda_{T}^{2}}{R_{{\rm TF}}^{2}2\pi} & =\int d\tilde{r}\tilde{r}f_{n}\left(\beta\mu-\frac{\tilde{r}^{2}}{T/T_{F}}\right).
\end{alignat}
Using the fact that $\lambda_{T}^{2}=2\pi\hbar/(Mk_{{\rm B}}T)$ and
$k_{{\rm B}}T_{F}=\hbar(2N\omega_{0}^{2})^{1/2}$, we have 
\begin{alignat}{1}
\frac{T_{F}}{4T}=\int d\tilde{r}\tilde{r}f_{n}\left(\beta\mu-\frac{\tilde{r}^{2}}{T/T_{F}}\right),
\end{alignat}
changing coordinates to $y=\tilde{r}^{2}/(T/T_{F})$ gives in total,
\begin{alignat}{1}
\frac{T}{T_{F}}=\left[2\int_{0}^{\infty}dyf_{n}\left(\beta\mu-y\right)\right]^{-1/2}
\end{alignat}
This definition allows us to find a reduced temperature for the trapped
system for a given $\beta\mu$ in the homogeneous system.

The spring constant 
\begin{alignat}{1}
\left(\nabla\rho_{0}\cdot\mathbf{u}_{n}\right)\left(\frac{\nabla V_{{\rm tr}}}{M}\cdot\mathbf{u}_{n}\right) & =\frac{\partial\rho_{0}}{\partial r}A_{n}\tilde{r}^{n+1}rA_{m}\omega_{\perp}\tilde{r}^{m+1}r\nonumber \\
\rho_{0}\left(\nabla\cdot\mathbf{u}_{n}\right)\left(\frac{\nabla V_{{\rm tr}}}{M}\cdot\mathbf{u}_{n}\right) & =\rho_{0}\frac{n+2}{R_{{\rm TF}}}\tilde{r}^{n}A_{n}\omega_{\perp}^{2}r\tilde{r}^{m+1}A_{m}\nonumber \\
\rho_{0}\left(\frac{\nabla V_{{\rm tr}}}{M}\cdot\mathbf{u}_{n}\right)\left(\nabla\cdot\mathbf{u}_{n}\right) & =\rho_{0}A_{n}\omega_{0}^{2}r\tilde{r}^{n+1}A_{n}\frac{m+2}{R_{{\rm TF}}}\tilde{r}^{m}\nonumber \\
c_{n}\left(\nabla\cdot\mathbf{u}_{n}\right)^{2} & =c_{n}\frac{n+2}{R_{{\rm TF}}}\frac{m+2}{R_{{\rm TF}}}r^{n}r^{m}A_{n}A_{m},\label{eq:sc4}
\end{alignat}

The spring constants cancel leaving only Eq.~\eqref{eq:sc4} to calculate,
\begin{alignat}{1}
K_{nm}= & (n+2)(m+2)\frac{M\omega_{0}^{2}}{2k_{{\rm B}}T_{{\rm F}}}\int d\mathbf{r}\left[n_{0}\left(\frac{\partial P}{\partial n}\right)_{\bar{s}}\right]\,\tilde{r}^{n+m},\nonumber \\
= & (n+2)(m+2)\frac{M\omega_{0}^{2}R_{F}^{2}}{2k_{{\rm B}}T_{{\rm F}}\lambda_{T}^{2}}k_{{\rm B}}T\nonumber \\
 & \times\int d\mathbf{r}f_{n}\left(\beta\mu-\frac{M\omega_{0}^{2}r^{2}}{2k_{{\rm B}}T}\right)\left(\frac{\partial P}{\partial n}\right)_{\bar{s}}\,\tilde{r}^{n+m},\nonumber \\
= & (n+2)(m+2)\frac{M\omega_{0}^{2}R_{F}^{2}\pi}{\lambda_{T}^{2}}\frac{T}{T_{{\rm F}}}\nonumber \\
 & \times\int_{0}^{\infty}drf_{n}\left(\beta\mu-\frac{r^{2}}{T/T_{{\rm F}}}\right)\left(\frac{\partial P}{\partial n}\right)_{\bar{s}}\,\tilde{r}^{n+m+1}.
\end{alignat}

\section{Solving for ${\rm {det}\mathbf{A}(\tilde{\omega})=0}$}

\label{App:solvedet}

In order to solve the matrix equation $\mathbf{A}(\tilde{\omega})\,\mathbf{x}=0$,
where the vector of displacement fields is, $\mathbf{x}=[A_{0},\dots,\,A_{n},\dots]^{{\rm {T}}}$,
we expand the matrix $\mathbf{A}(\tilde{\omega})=\mathbf{M}\tilde{\omega}^{2}-\mathbf{K}$
to find the eigenvalues. Here, $\mathbf{M}$ and $\mathbf{K}$ are
the matrices of the reduced weighted mass moments and the spring constants,
respectively. The matrix $\mathbf{M}$ can be written as a product
of a lower triangular matrix $\mathbf{L}$, its conjugate transpose,
and a diagonal matrix $\mathbf{D}$, 
\begin{alignat}{1}
\mathbf{M}=\mathbf{L}\cdot\mathbf{D}\cdot\mathbf{L}^{{\rm {T}}}.
\end{alignat}
In terms of this decomposition, the matrix equation $\mathbf{A}(\tilde{\omega})\,\mathbf{x}=0$
is written as, 
\begin{alignat}{1}
\left[\mathbf{D}^{-1}\cdot\mathbf{L}^{-1}\cdot\mathbf{K}\cdot(\mathbf{L}^{-1})^{{\rm {T}}}\right]\mathbf{L}^{{\rm {T}}}\,\mathbf{x}=\tilde{\omega}^{2}\mathbf{L}^{{\rm {T}}}\,\mathbf{x}.
\end{alignat}
We see that the eigenvalues of the positive definite matrix, $\left[\mathbf{D}^{-1}\cdot\mathbf{L}^{-1}\cdot\mathbf{K}\cdot(\mathbf{L}^{-1})^{{\rm {T}}}\right]$
are the mode frequencies, $\tilde{\omega}$. The displacement field
for each eigenvalue can also be calculated similarly, with each eigenstate
of the matrix $\left[\mathbf{D}^{-1}\cdot\mathbf{L}^{-1}\cdot\mathbf{K}\cdot(\mathbf{L}^{-1})^{{\rm {T}}}\right]$.
Once the displace fields for a mode are found, we may calculate its
density fluctuation.

\bibliographystyle{apsrev4-1}
\bibliography{coll_modes}

\end{document}